# Fabrication of nanometer-spaced electrodes using gold nanoparticles

Saiful I. Khondaker and Zhen Yao[a]

*Department of Physics, Texas Materials Institute, and Center for Nano- and Molecular Science and Technology, The University of Texas at Austin, Austin, TX 78712*

A simple and highly reproducible technique is demonstrated for the fabrication of metallic electrodes with nanometer separation. Commercially available bare gold colloidal nanoparticles are first trapped between prefabricated large-separation electrodes to form a low-resistance bridge by an ac electric field. A large dc voltage is then applied to break the bridge via electromigration at room temperature, which consistently produces gaps in the sub-10 nm range. The technique is readily applied to prefabricated electrodes with separation up to 1 μm, which can be defined using optical lithography. The simple fabrication scheme will facilitate electronic transport studies of individual nanostructures made by chemical synthesis. As an example, measurement of a thiol-coated gold nanoparticle showing a clear Coulomb staircase is presented.

Nanostructures made by chemical synthesis such as nanocrystals and molecules have attracted tremendous attention as they constitute promising building blocks for future generation of electronic devices and model systems to investigate novel quantum transport phenomena in confined systems. Probing the electronic properties of individual nanostructures requires the placement of nanostructures between electrodes with nanometer separation. Scanning tunneling microscopy has proven to be a useful approach.[1-3] It is, however, more advantageous to employ nanofabricated electrodes in a fixed planar geometry as they offer greater mechanical stability and allow for straightforward implementation of electrostatic gates. Standard electron-beam lithography (EBL) has a typical resolution of >20 nm, which is not sufficient to define electrodes to contact most chemical nanostructures of interest. Several techniques have been demonstrated to be able to produce gaps in the sub-10 nm range, which involve either highly elaborate EBL[4] or special techniques and processes such as mechanical break junctions,[5] shadow evaporation,[6] metal deposition over suspended structures,[7] electroplating,[8] selective etching of molecular-beam epitaxy structures,[9] and electromigration-induced breaking of thin metal wires defined by EBL.[10] The complexity of these techniques and/or the need to access high-resolution lithography, however, have greatly limited their general applications.

In this letter, we present a remarkably simple and highly reproducible method for the fabrication of nanometer-spaced metallic electrodes starting from prefabricated leads of up to 1 μm separation, which can easily be defined using standard optical lithography. The fabrication is done in two steps. Commercially available bare gold colloidal nanoparticles are first assembled between the prefabricated electrodes to form a low resistance bridge by an ac electric field. A dc bias voltage is then ramped across the bridge at room temperature until it is broken due to current-induced electromigration. We find that this process consistently produces stable electrodes of sub-10 nm separation. Our technique does not require high-resolution lithography and other special instruments, and therefore, is expected to greatly facilitate the connection of individual nanometer-sized structures to external electrical measurement setup. Here we show an example of transport characteristics of a thiol-coated gold nanoparticle.

The starting electrodes used in this work are fabricated on oxide-coated silicon substrates using standard EBL and lift-off of thermally-evaporated metal films consisting of 5 nm of chromium and 30 nm of gold. The initial separation between the electrodes are in the range of 40 nm to 1 μm. Trapping of gold nanoparticles between the electrodes is done in a probe station following the work of Amlani *et al*.[11] We use commercial gold colloidal nanoparticles in sizes of 2, 5, 10, 20, and 50 nm suspended in water.[12] Prior to trapping, the electrodes are cleaned in an $O_2$ plasma for ~30 min and rinsed in ethanol. Figure 1(a) shows a schematic diagram of our experimental setup. A small drop of gold nanoparticle solution is placed on the substrate to cover the electrodes and an ac bias with a frequency of 1 MHz is applied between the electrodes using a function generator. Due to the presence of a field gradient, the nanoparticles experience a dielectrophoretic force that attracts them to the gap where the electric field gradient is the largest. An output voltage is observed in the oscilloscope as soon as the particles have bridged the gap, at which point trapping is self-terminated. The circuit is then disconnected and the sample is dried with a $N_2$ gas flow. We find that the time and voltage required for the trapping process increase with the initial size of the gaps: they are typically ~1-3 s and ~1-2 V for gaps below 100 nm, and ~1 min and ~5 V for 1 μm gaps.

Figure 1(b) shows a scanning electron microscope (SEM) image of a single ~50 nm Au particle trapped in a 45 nm gap. The two-terminal resistance of the nanoparticle bridge is 332 Ω. We find that individual particles are trapped in the gaps in more than 50% cases if the initial gaps are smaller than the sizes of the nanoparticles. The nanometer-sized gap is created via current-induced electromigration[10] by slowly ramping a dc voltage across the nanoparticle bridge. Figure 1(c) shows the current as a function of bias voltage. At 2.25 V, the current abruptly drops to essentially zero signaling the breaking of the bridge. SEM image taken afterwards shows a gap below 10 nm as indicated by an arrow [Fig. 1(d)]. Figure 1(e) displays a field-emission SEM image of another pair of prefabricated electrodes of 400 nm separation bridged with 50 nm Au particles. Interestingly, some nanoparticles appear to have been fused together. SEM image after the breaking process again shows a gap below 10 nm [Fig. 1(f)].

---

[a] Electronic mail: yao@physics.utexas.edu

Similar experiments have been performed for nanoparticles with sizes of 2-50 nm and for electrodes with initial gap sizes of 40 nm to 1 µm. The resistances of our nanoparticle bridges (including those consisting of a chain of many particles) are consistently in the range of 300 to 650 Ω, which are at least one order of magnitude lower than those reported by Amlani et al.[11] We find that if a smaller ac amplitude is employed, nanoparticles can still bridge the gap, but the resistance of the bridge is much higher; in this case, no voltage output can be detected in the oscilloscope. The low resistances of our nanoparticle bridges are the key that allows us to apply the breaking technique to create nanometer gaps, and moreover, suggest that the nanoparticle chains can be used as ideal electrical leads for contacting other nanometer-sized objects.

In approximately 50% of the samples that we have studied, including 40 pairs of electrodes with 200 nm to 1 µm initial separation and 40 pairs of electrodes with 40-100 nm initial separation, we find that the resulting gaps to be less than the resolution limit (~4 nm) of our field-emission SEM (LEO-1530). We also note that when nanoparticles of smaller sizes are used, the breaking typically occurs at lower voltages, which in turn yields smaller gap sizes. The nanometer-sized gaps formed by the nanoparticles are extremely robust mechanically. SEM inspection of the samples after ultrasonic agitation up to 30 minutes does not show any sign of damage to the gaps.

The major advantage of our technique is that it does not require EBL and other special instrumentation. The initial electrodes with ~1 µm gap can easily be fabricated using optical lithography. In previous work employing electromigration to create gaps in metallic nanowires defined by EBL, cryogenic temperatures are required to break the wires in order to create nanometer-sized gaps.[10,13,14] In our new technique, however, this can be simply achieved at room temperature, presumably because there tend to be more weak links in our nanoparticle bridges which facilitate the breaking.

The nanometer-spaced electrodes we have created are ideally suitable to connect individual chemically synthesized nanostructures. Here we show preliminary measurement of a thiol-coated 2-3 nm Au nanoparticle[15]. The thiol coating, which plays a dual role of both a linker that binds the nanoparticle to electrodes and a tunnel barrier, is formed by adding dithiothreitol ($C_4H_{10}O_2S_2$), a water soluble thiol, into a water suspension containing bare Au nanoparticles and waiting for ~24 h. An ac electric field is again employed to trap these thiol-coated nanoparticles. Unlike bare nanoparticles, however, there is a high tunnel resistance between the thiol-coated nanoparticles and the electrodes, which prevents us from observing the output voltage when the particles bridge the gap. The probability of trapping individual nanoparticles is optimized by tuning the density of the nanoparticle solution and the trapping time.

Figure 2 shows a plot of current $I$ as a function of bias voltage $V$ for one of the devices measured at 4.2 K, which exhibits a suppression of current around zero bias as well as current steps with approximately constant spacing at large bias. The behavior seems to be consistent with the Coulomb staircase phenomenon which occurs in a double-barrier single-electron tunneling system with asymmetric barriers.[16] The dotted line in Fig. 2 shows a Monte Carlo simulation[17] based on the orthodox theory of single-electron tunneling. The capacitances and resistances for the two tunnel barriers used in the simulation are $C_1$=0.89 aF, $C_2$=0.73 aF, $R_1$=50 MΩ, and $R_2$=3.7 GΩ. In addition, there is a background offset charge of $Q_0$ = 0.11 e. At large bias voltages, there is some deviation in the magnitude of the current between the simulation and experiment. This may be attributed to an effective tunnel barrier that depends on the bias voltage.[7,18] The Coulomb charging energy of the nanoparticle as determined from the fitting parameters is $E_C$= $e^2/(C_1+C_2)$ =98 meV, which is consistent with a simple estimate of the capacitance of a 3 nm diameter metal sphere at a distance of 0.75 nm from two parallel metal planes in a dielectric environment with ε=2.8.[19]

In conclusion, we have developed a simple technique to create electrodes with nanometer separation, which does not require high-resolution lithography, by combining electrostatic trapping of commercial gold colloidal nanoparticles and current-induced electromigration. The technique is expected to find extensive general use in the electrical characterization of a variety of individual chemical nanostructures. Our preliminary measurement of a thiol-coated 3 nm Au nanoparticle shows a Coulomb staircase behavior.

We thank D. J. Solis and A. M. Belcher for suggesting the type of thiol molecules used in this work, D.-H. Chae for experimental assistance, and the Welch Foundation and the Alfred P. Sloan Foundation for partial financial support.

**FIGURE CAPTIONS**

FIG. 1. Electrostatic trapping of nanoparticles between prefabricated electrodes and formation of nanometer gaps by current-induced electromigration. (a) is a schematic diagram of the experimental setup for nanoparticle trapping. (b) is a SEM image of a 50 nm Au nanoparticle bridging a pair of electrodes with 45 nm gap. The nanoparticle bridge is broken by applying a dc voltage, as shown in the current-voltage trace in (c). SEM image taken afterwards is shown in (d) where the resulting gap is indicated by an arrow. Another example is shown in (e) displaying a field-emission SEM image of a pair of prefabricated electrodes with 400 nm gap bridged by a chain of 50 nm Au nanoparticle and (f) the resulting sub-10 nm gap after breaking. The scale bars in the images correspond to 200 nm.

FIG. 2. *I-V* characteristic of a thiol-coated 3 nm Au nanoparticle measured at 4.2 K. The solid line represents the experimental data whereas the dashed line is a fit using the orthodox Coulomb blockade model. The curves are offset for clarity.

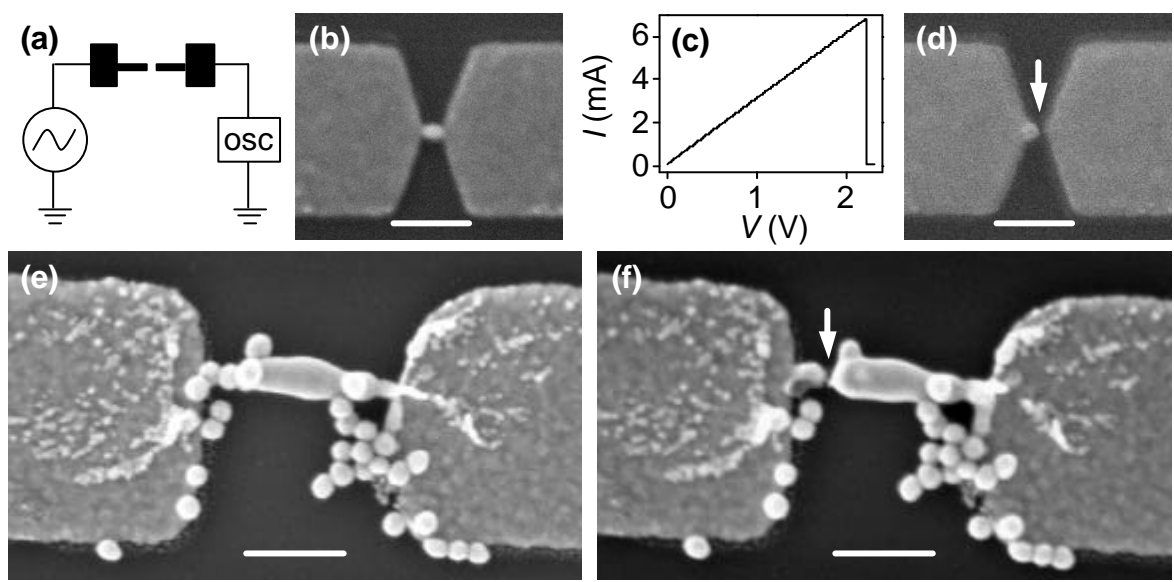

FIG. 1 Khondaker *et al.*

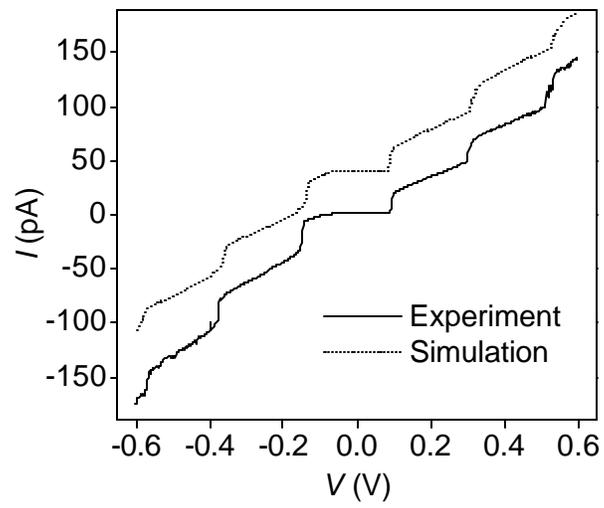

**FIG. 2 Khondaker *et al.***